\documentclass[aps,prl,showpacs,amsmath,twocolumn,amssymb,superscriptaddress,letterpaper]{revtex4}
\usepackage{times}
\usepackage{amsfonts}
\usepackage{mathrsfs}
\usepackage{graphicx}% Include figure files
\usepackage{dcolumn}% Align table columns on decimal point
\usepackage{bm}% bold math
\usepackage{color}

\usepackage[colorlinks,bookmarks=false,citecolor=blue,linkcolor=red,urlcolor=blue]{hyperref}
\def\be{\begin{equation}}       \def\ee{\end{equation}}
\def\bea{\begin{eqnarray}}      \def\eea{\end{eqnarray}}

\begin{document}
\title{ Topological Characters in Fe(Te$_{1-x}$Se$_x$) thin films}

\author{Xianxin Wu}
\affiliation{ Institute of Physics, Chinese Academy of Sciences,
Beijing 100190, China}

\author{Shengshan Qin}
\affiliation{ Institute of Physics, Chinese Academy of Sciences,
Beijing 100190, China}

\author{Yi Liang}
\affiliation{ Institute of Physics, Chinese Academy of Sciences,
Beijing 100190, China}

\author{Heng Fan }  \affiliation{ Institute of Physics, Chinese Academy of Sciences,
Beijing 100190, China}
\affiliation{Collaborative Innovation Center of Quantum Matter, Beijing, China}

\author{Jiangping Hu  }\email{jphu@iphy.ac.cn} \affiliation{ Institute of Physics, Chinese Academy of Sciences,
Beijing 100190, China}\affiliation{Department of Physics, Purdue University, West Lafayette, Indiana 47907, USA}
\affiliation{Collaborative Innovation Center of Quantum Matter, Beijing, China}

\date{\today}

\begin{abstract}
We investigate topological properties in the Fe(Te,Se) thin films. We find that the single layer FeTe$_{1-x}$Se$_x$ has nontrivial $Z_2$ topological invariance which originates from the parity exchange at $\Gamma$ point of Brillouin zone. The nontrivial topology is mainly controlled by the Te(Se) height.   Adjusting the height, which can be realized as function of $x$ in  FeTe$_{1-x}$Se$_x$,   can drive a topological phase transition.  In a bulk material,  the two dimensional $Z_2$ topology invariance is extended to a strong  three-dimensional one. In a thin film, we predict that the topological invariance oscillates with the number of layers. The results can also be applied to iron-pnictides. Our research establishes FeTe$_{1-x}$Se$_x$ as a unique system to integrate high T$_c$ superconductivity and topological properties  in a single electronic structure.
\end{abstract}


\maketitle
Recently, topological insulators(TI)\cite{Kane2005,Hasan2010,Qi2011} and iron based superconductors\cite{Kamihara2008,Johnston2010} have attracted enormous attention in condensed matter physics. Most topological insulators are semiconductors with strong spin orbital coupling in which the bulk gap protects  surface or edge states depending on dimensionality. Their band structures are attributed to $p$-orbitals.  Iron-based superconductors, including  iron-pnictides and iron-chalcogenides, are  multi-orbital electronic systems in which the electronic structures are mainly attributed  to Fe $3d$ orbitals. Therefore,  these two systems appear to be distantly apart. However, the combination of topological insulators and superconductors is known to generate novel physics, for example, Majorana fermions\cite{Fu2010,Linder2010}. The usual way  to integrate TI and superconductivity is through superconducting proximity effect by making TI-superconductor heterostructures.  Iron-based superconductors, due to their short coherent  length and material incompatibility, are not suitable to such an integration.

However, recently,  pioneer studies suggest that iron-based superconductors  can carry intrinsic nontrivial topological properties. Two  examples have been provided. One is the single layer FeSe grown on SrTiO$_3$ substrate where  nontrivial $Z_2$ topology can be tuned through the band inversion at $M$ point in Brillouin zone \cite{Hao2014}. The other is CaFeAs$_2$ where nontrivial topology exists on additional As layers that are inserted between FeAs layers\cite{Wu2014}. Although these results are encouraging,  in the first case, a fine-tuning is required and in the second case, the topological properties are not from the layers that are responsible for high temperature superconductivity.

In this paper,  we report  that  a single layer FeTe$_{1-x}$Se$_{x}$ with $x$  less than a critical value $x_c$ which is estimated to around 0.7,  carries nontrivial $Z_2$ topological invariance  that is originated from the parity exchange at $\Gamma$ point.  We identify that  the nontrivial topology is mainly controlled by the Te(Se) height.   Adjusting the height, which can be realized as function of $x$ in  FeTe$_{1-x}$Se$_x$,   can drive a topological phase transition. An effective model  is constructed to explicitly describe the topological physics. In the bulk material, the two dimensional $Z_2$ topology is extended into a strong  three-dimensional $Z_2$ invariance.  In a thin film,   it  exhibits oscillation behavior  with a trilayer structure being topologically trivial.  The results can also be applied to iron-pnictides.  These results establish FeTe$_{1-x}$Se$_x$ as a unique system to integrate high T$_c$ superconductivity and topological properties  in a single electronic structure.

\begin{figure}[t]
\centerline{\includegraphics[height=5 cm]{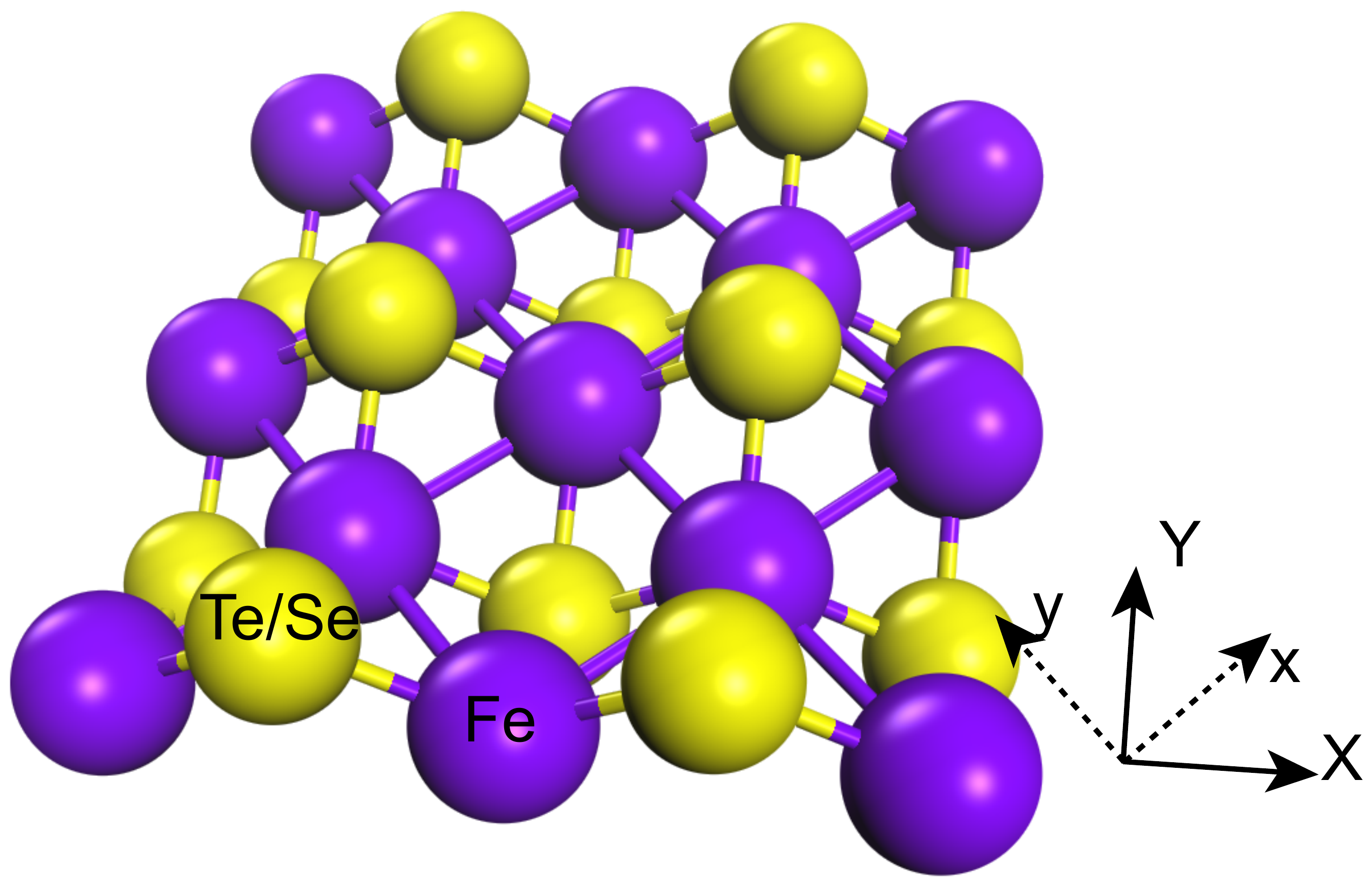}}
\caption{(color online). Schematic view of the structure of FeTe(FeSe).
 \label{structure} }
\end{figure}

\begin{figure}[t]
\centerline{\includegraphics[height=4.5 cm]{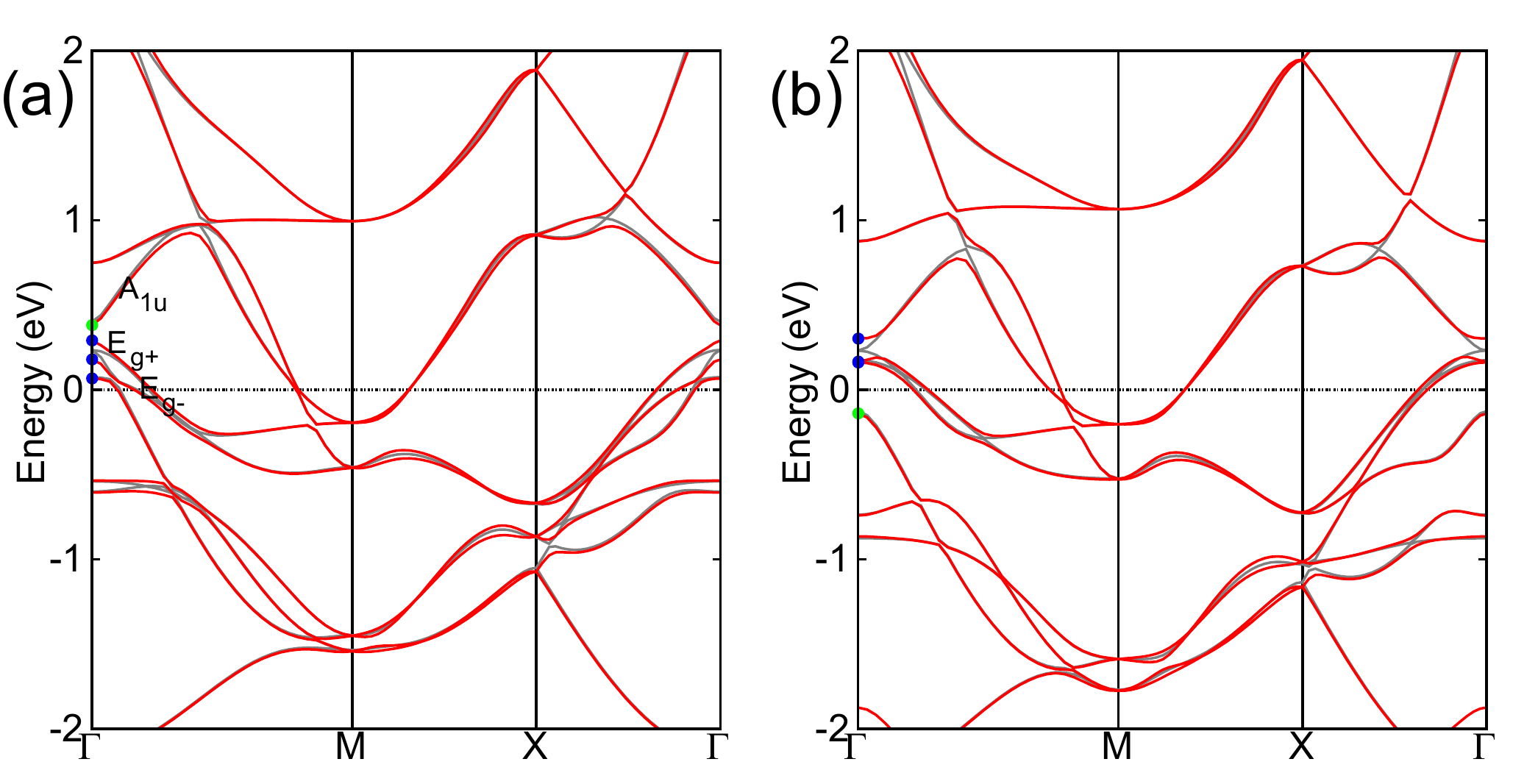}}
\caption{(color online). Band structures for monolayer FeTe: (a) $a=3.925$\AA~ and (b) $a=3.805$ \AA.~ The red solid lines represent the band with SOC and the gray lines the band without SOC. The parities of the eigenstates at $\Gamma$ near the Fermi level are shown : blue circles for even parities and green circles for odd parities.
 \label{bandsoc} }
\end{figure}

{\it $Z_2$ topological invariance in the band structure of a single layer FeTe: } The crystal structure of FeTe (FeSe) is shown in Fig.\ref{structure}. In order to gain the insight of the nontrivial topological properties,  first, we analyze a topologically trivial case, the band structure of monolayer  FeTe with  lattice constant $a$= 3.925\AA, where the lattice constant is close to that of monolayer FeSe on SrTiO$_3$\cite{Liu2012Lu,Bazhirov2013,Zheng2013_Wang}. In Fig.\ref{bandsoc}(a),  the  band structure of monolayer FeTe  is plotted.   Similar to other iron based superconductors, there are both hole pockets at $\Gamma$ and electron pockets at $M$.  We consider the effect of spin orbital coupling (SOC). Comparing the bands with SOC (red lines) and without SOC (gray lines), we find that  SOC has two main effects: (1) the two-fold degenerate $E_g$ bands at $\Gamma$ are split into $E_{g+}$ and $E_{g-}$; (2) the Dirac cone in the $\Gamma-M$ line is also gapped. The latter makes a local gap  at every $k$ point in Brillouin zone, which is important for defining the topological properties discussed in the following. The parities of the eigenstates at $\Gamma$ near the Fermi level are shown in Fig.\ref{bandsoc}(a). The odd parity state $A_{1u}$ is contributed by Fe $d_{xy}$ orbitals strongly coupled with Te $p_z$ orbitals. The two even parity states $E_{g+}$ and $E_{g-}$ are attributed to Fe $d_{xz}$, $d_{yz}$ strongly hybridized with Te $p_x$, $p_y$ orbitals while the other even parity state is mainly attributed to $d_{xy}$ orbitals. Even through SOC is relatively weak in iron, the separation between $E_{g+}$ and $E_{g-}$ is large because the $E_g$ states involve both Fe $3d$ and Te $5p$ states. With SOC, each band at $X$ and $M$ is four-fold degenerate and contains two Kramers pairs, which are protected by the crystal symmetry. Each Kramers pair has a definite parity. The two Kramers pairs at $M$ are related to each other by the mirror reflection $\hat{g}_1=\{\sigma^{X}|00\}$.  The two Kramers pairs at $X$  are governed by the glide plane $\hat{g}_2=\{\sigma^{Z}|\frac{1}{2}\frac{1}{2}\}$\cite{Cvetkovic2013}. As the inversion operator $\hat{I}=\{i|\frac{1}{2}\frac{1}{2}\}$ anti-commutate with $\hat{g}_1$ at $M$ and $\hat{g}_2$ at $X$, the two Kramers pairs at $M$ and $X$ have opposite parities. Therefore, the parity product of two Kramers pairs in each four-fold degenerate state at X and M is negative. The product of parities of occupied bands at X and M can be obtained easily by just counting the number of the occupied bands. The band structure in Fig.\ref{bandsoc}(a) is  topologically trivial according to the calculation of parity product of occupied states at the four time-reversal-invariant momenta\cite{Fu2007}.

{\it Anion height  and topological phase transition: } From the above symmetry analysis, one can conclude that  band inversion at M or X will not change the $Z_2$ topological number of the system. However, the situation is different at $\Gamma$.
%Only band inversions at $\Gamma$ point can drive the system into a $Z_2$ topologically nontrivial phase.
One can notice that  the two states $A_{1u}$ and $E_{g+}$ at $\Gamma$ , which have opposite parities, are separated by a small gap about 0.09 eV. If the gap  can  be closed and reopened, the inversion between the two states will drive the system into a $Z_2$ topologically nontrivial phase because of the change of the overall parity character.

\begin{figure}[t]
\centerline{\includegraphics[height=4 cm]{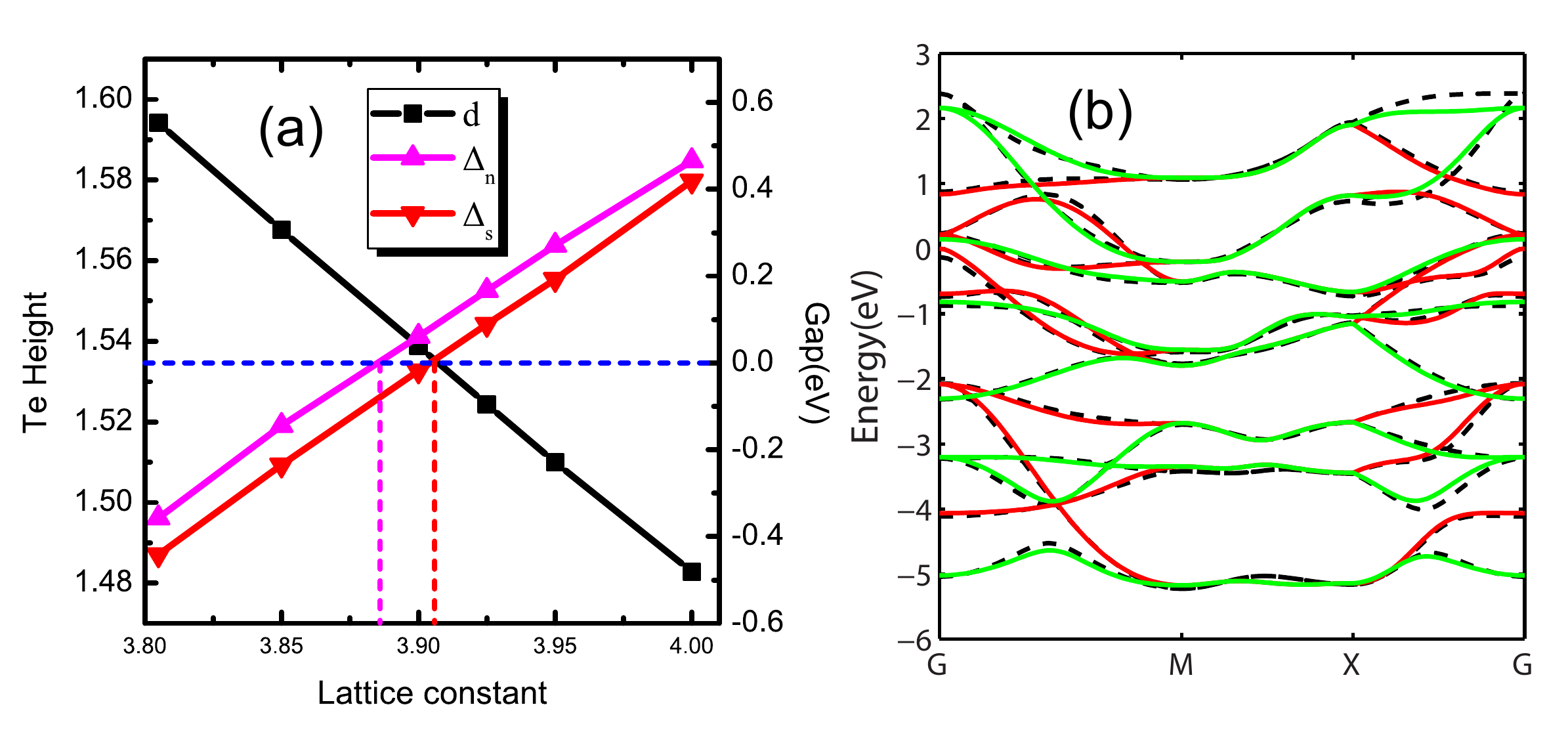}}
\caption{(color online). (a) The Te height $d$ and the gaps $\Delta_n$ and $\Delta_s$ as function of the inplane lattice constant. $\Delta_s$ and $\Delta_n$ are the energy differences between $A_{1u}$ and $E_{g+}$ states with SOC and without SOC. Both Te height and lattice constant are given in Angstrom. (b) Band structures of monolayer FeTe without SOC from tight bind model and DFT. The red lines represent $k$ bands and the green lines represent $k+Q$ bands. The black dash lines represent the DFT bands.
 \label{phasebandtb} }
\end{figure}

We find that this inversion takes place in  the monolayer FeTe with a fully relaxed lattice constant parameter $a=3.805$ \AA.~ In the Fig.\ref{bandsoc}(b), we plot the band structure of the monolayer FeTe, where the two states $A_{1u}$ and $E_{g+}$ at $\Gamma$ are clearly inverted.  We further investigate the origin of this band inversion and  find that it is controlled by the Te height  and lattice constants. We define $\Delta_s$ and $\Delta_n$ as the energy differences between $A_{1u}$ and $E_{g+}$ states with SOC and without SOC, respectively. The Te height $d$, $\Delta_s$ and $\Delta_n$ as functions of lattice constants are shown in Fig.~\ref{phasebandtb}(a). As the decreasing of lattice constant, the $A_{1u}$ state sinks while the $E_{g+}$ state rise, resulting the decreasing of $\Delta_s$ and $\Delta_n$.  Our calculations with SOC show that the phase transition occurs at $a=3.905$\AA, where the corresponding Te height is $d=1.535$\AA. For $a<3.905$\AA,~ the system becomes topologically nontrivial. When $3.886<a<3.905$, $\Delta_s$ is negative but $\Delta_n$ is positive, indicating that the band inversion is completely driven by SOC. This mechanism is similar to that in 3D topological insulator Bi$_2$Se$_3$\cite{Zhang2009}. $\Delta_n$ is negative for $a<3.886$\AA ~and band inversion has happened without SOC, leaving the two $E_g$ bands quadratic touching, which is the case shown in Fig.\ref{phasebandtb}(b).

As the anions height can also be tuned by the concentration of Se in FeTe$_{1-x}$Se$_x$, the topological phase transition can take place  by changing the Se concentration.  To estimate the critical concentration, $x_c$,  we take the experimental data of the lattice constants. The Se heights in FeSe and FeTe$_{0.5}$Se$_{0.5}$ are 1.46 \AA~and 1.589 \AA\cite{Lehman2010}, respectively. The topological phase transition point is roughly at $x_c=0.7$ by assuming that the anion height is linear with respect to the Se concentration in FeTe$_{1-x}$Se$_x$.
%The relaxed lattice constant of monolayer FeTe is 3.805 \AA and its band structure is shown in Fig.\ref{bandsoc}(b).

{\it Tight binding model and effective model: }
%\subsection{tight binding model }
To understand the parity exchange at $\Gamma$, we start from a tight binding model of FeTe system. As Fe $d$ orbitals hybridize strongly with Te $p$ orbitals, an effective model containing only Fe $d$ orbtials is not good enough to describe the system.  A general model including Fe $d$ and X (X=Te,Se) $p$ oribtals for iron based superconductors can be written as,
\begin{eqnarray}
H_t=\sum_{\alpha\beta}\sum_{mn}\sum_{ij}(t^{mn}_{\alpha\beta,ij}+\epsilon_{\alpha}\delta_{mn}\delta_{\alpha\beta}\delta_{ij})c^{\dag}_{\alpha m\sigma}(i)c_{\beta n \sigma}(j)
\end{eqnarray}
Here, $\alpha$,$\beta$ labels the sublattices (A and B for Fe and Te). $\sigma$ labels the spin and $m$,$n$ label the $d$ and $p$ orbitals. $i$,$j$ label lattice sites. $t^{mn}_{\alpha\beta,ij}$ are  the corresponding hopping parameters. $\epsilon_{m}$ are the onsite energies of $d$ or $p$ orbitals. $c^{\dag}_{\alpha m \sigma}(i)$ creates a spin-$\sigma$ electron in $m$ orbital of $\alpha$ sublattice at site $i$. We can eliminate the sublattice index by  writing the Hamiltonian in momentum space with respect to one Fe unit cell,
\begin{eqnarray}
H_t=\sum_{k\in BZ1}\phi^{\dag}(k)A(k)\phi(k)
\end{eqnarray}
%Here, $k$ is defined in the Brillioun Zone of one Fe unit cell (BZ1). $\phi(k)=[c_{xz}(k),c_{yz}(k),c_{x^2-y^2}(k+Q),c_{xy}(k+Q),c_{z^2}(k+Q),c_{x}(k+Q),c_{y}(k+Q),c_{z}(k)]^T$ with $Q=(\pi,\pi)$ defined in BZ1.  The connection between the bases in one Fe unit cell and two Fe unit cell is,
%\begin{eqnarray}
%c_{\alpha}(k)&=&\frac{\sqrt{2}}{2}(c_{A\alpha}(q)+c_{B\alpha}(q)),\\
%c_{\alpha}(k+Q)&=&\frac{\sqrt{2}}{2}(c_{A\alpha}(q)-c_{B\alpha}(q)),
%\end{eqnarray}
%where $q$ is defined in BZ of two Fe unit cell (BZ2).
In this case, as the inversion symmetry exchange the two sublattices, at $\Gamma$ and $M$ points the parity of $c_{m}(0)$ is even and $c_{m}(Q)$ odd for $\alpha$ being $d$-orbital  while the parity of $c_{m}(0)$ is odd and $c_{m}(Q)$ even for $\alpha$ being $p$ orbital\cite{Hu2013,Hu2012}.

\begin{figure}[t]
\centerline{\includegraphics[height=2.7 cm]{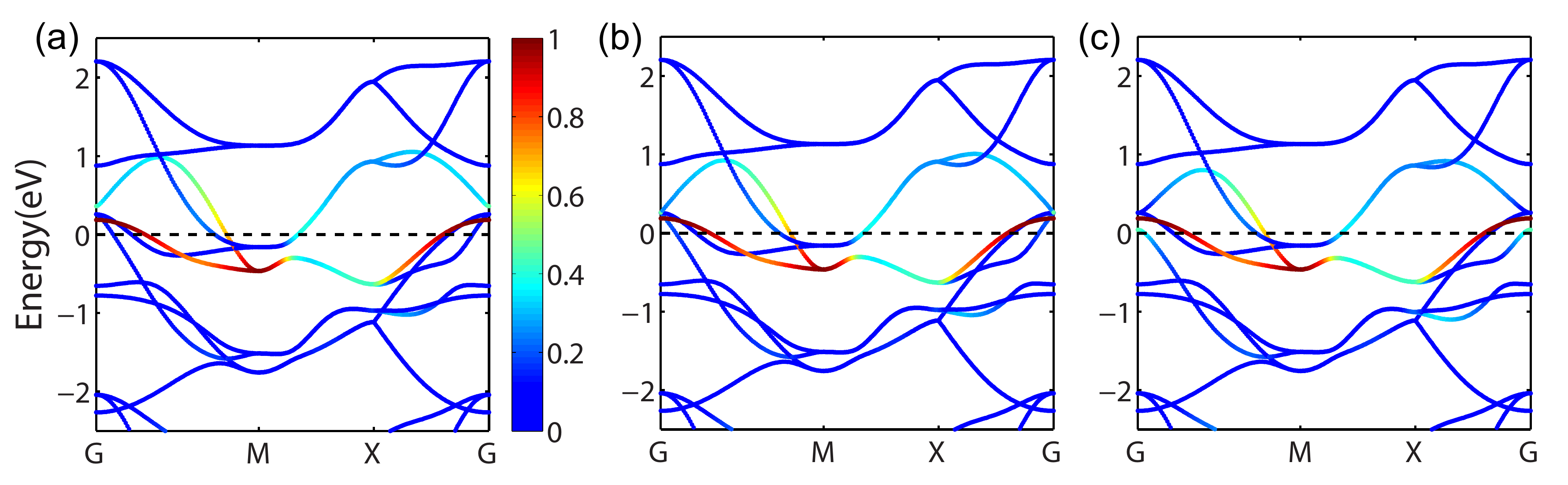}}
\caption{(color online). The band structures with different hopping parameter $q^{43}_{xy}$. (a) $q^{43}_{xy}=0.649$ eV. (b) $q^{43}_{xy}=0.624$ eV. (c) $q^{43}_{xy}=0.569$ eV. The color represents the weights of Fe $d_{xy}$ orbitals.
 \label{bandtbinvert} }
\end{figure}

  Without SOC, the $H_t$  can be constructed, as shown in Fig.\ref{phasebandtb}(b), by  fitting the DFT band structure. The red bands are referred as ``$k$'' bands and the green ones are referred as ``$k+Q$" bands.
We find that the inversion is mainly controlled by the  coupling between Fe $d$ orbital and Te $p$ orbital, which is described by the hopping parameter $q^{43}_{xy}$, the nearest neighbor (NN) hopping between Fe $d_{xy}$ and Te $p_z$ as specified in supplement materials. As the increasing of lattice constant, the height of Te in monolayer FeTe decreases and the coupling between Fe $d$ orbital and Te $p$ orbital becomes stronger so that $q^{43}_{xy}$  changes significantly.  In Fig.\ref{bandtbinvert}(a)-(c), we plot the band structures with different $q^{43}_{xy}$. As the decreasing of $q^{43}_{xy}$, the odd parity band, attributed to Fe $d_{xy}$ and Te $p_z$ orbitals, sinks below the even parity bands.

How to understand this effect of $q^{43}_{xy}$? From the tight binding Hamiltonian, we know that the NN coupling term between Fe $d_{xy}$ and Te $p_z$ is $4q^{43}_{xy} cos(k_x/2)cos(k_y/2)$. Thus, at $\Gamma$ point this coupling vanishes in the ``$k+Q$" bands. However, it is maximized at $\Gamma$ in the "$k$" bands.  The energy difference between the two states is proportional to the coupling parameter. Therefore, the top state at $\Gamma$ sinks with the decreasing of $q^{43}_{xy}$. When the gap between $A_{1u}$ and $E_g$ states is less than 80 meV, SOC can induce band inversion. When the $A_{1u}$ state sinks below the $E_g$ state,  away from $\Gamma$ point $d_{xy}$ orbital can couple with $d_{xz}$ and $d_{yz}$ orbitals, which produces an anti-crossing. After the parity exchange, shown in Fig.\ref{bandtbinvert}(c), the system becomes topologically nontrivial with SOC.

An effective Hamiltonian can be derived to capture  the parity exchange bands near $\Gamma$ point by including SOC.  The detailed  construction is provided in the supplement materials. Here we briefly summarize the result.  With SOC, the total Hamiltonian can be written as $H=H_t+H_{sf}+H_{snf}$, where $H_{sf}$ and $H_{snf}$ describe spin-flip and spin-nonflip SOC terms respectively. Focusing on $\Gamma$ point,  $d_{xz}$ and $d_{yz}$ orbitials couple with $p_x$ and $p_y$ orbitals and $d_{xy}$ orbital couples with $p_z$ orbital. We combine them to form new bases and the details are given in  the supplementary.
In the basis $\Psi_{eff}(k)=[c_{1+,\uparrow}(k),c_{-1+,\uparrow}(k),c_{0+,\uparrow}(k),c_{-1+,\downarrow}(k),c_{1+,\downarrow}(k),c_{0+,\downarrow}(k)]$, we can get the effective model around $\Gamma$ point as,
\begin{eqnarray}
&&H_{eff}=\sum_k \Psi_{eff}^{\dag}(k)h_{eff}(k)\Psi_{eff}(k),\\
&&h_{eff}(k)=\left(
\begin{array}{cc}
h_1(k) & 0\\
0 & h^{\star}_1(-k)\\
\end{array}
\right).
\end{eqnarray}
%The $h(k)$ has the following form,
%\begin{widetext}
%\begin{eqnarray}
%h_1(k)=\left(
%\begin{array}{ccc}
%A_1(k^2_x+k^2_y)+C_1 & A_2(k^2_x-k^2_y)+iB_2 k_xk_y & A_3(ik_x-k_y)\\
%A_2(k^2_x-k^2_y)-iB_2 k_xk_y &A_4(k^2_x+k^2_y)+C_4 & A_5(-ik_x-k_y)\\
%A_3(-ik_x-k_y) & A_5(ik_x-k_y) & A_6(k^2_x+k^2_y)+C_6 \\
%\end{array}
%\right).
%\end{eqnarray}
%\end{widetext}
The $h_1(k)$ is a $3\times 3$ matrix given in the supplementary. The bands of this effective model can  capture the DFT bands around $\Gamma$ point(see supplementary material).  As the band energy  attributed to the $c_{-1+,\uparrow}(k)$ is typically much lower, the six band structure can be further approximated by  a minimum effective four band structure in the new basis $\tilde{\Psi}_{eff}(k)=[c_{1+,\uparrow}(k),c_{0+,\uparrow}(k),c_{-1+,\downarrow}(k),c_{0+,\downarrow}(k)]$, the final effective Hamiltonian reads,
\begin{eqnarray}
H_{eff}&=&\sum_k \tilde{\Psi}_{eff}^{\dag}(k)\tilde{h}_{eff}(k)\tilde{\Psi}_{eff}(k),\\
\tilde{h}_{eff}(k)&=&\epsilon_0(k)+\left(
\begin{array}{cccc}
-M(k) & Ak_{+} & 0 & 0\\
Ak_{-} & M(k) & 0 & 0\\
0 & 0 & -M(k) & -Ak_{-}\\
0 & 0 & -Ak_{+} & M(k)\\
\end{array}
\right),
\end{eqnarray}
where $\epsilon_0(k)=C-D(k^2_x+k^2_y)$, $M(k)=M-B(k^2_x+k^2_y)$ and $k_{\pm}=k_x\pm ik_y$. $M<0$ corresponds to the inverted regime whereas $M>0$ corresponds to the normal regime, namely, the  topologically nontrivial regime. The fitting parameters for different lattice constants are given in Table \ref{4band}. The bands of this model are given in  the supplementary.

%\begin{table}[bt]%The best place to locate the table environment is directly after its first reference in text
%\caption{\label{formationE}%
%Parameters in the effective model for monolayer FeTe.}
%%%%% new table
%\begin{ruledtabular}
%\begin{tabular}{ccccccccccc}
%%\textrm{Left\footnote{Note a.}}& \textrm{Centered\footnote{Note
%%b.}}& \multicolumn{1}{c}{\textrm{Decimal}}&
%%\textrm{Right}\\
%%&\multicolumn{3}{c}{GaSe} & \multicolumn{3}{c}{GaS}\\
%lattice(\AA) & $C_1$  & $C_3$  & $C_4$  & $D_1$  & $D_2$  & $B_2$ & $D_3$ & $D_4$ & $A_1$ & $A_2$ \\
% \colrule
%3.805 & 0.305 & 0.170 & -0.146 & 0.131 & -4.881 & 11.118 & 2.601 & -8.439 & -0.250 & 0.696 \\
%3.9 & 2.97 & 5.72 & 1.53 & 4.29 & 7.35 & 2.33  & & & &\\
%\end{tabular}
%\end{ruledtabular}
%\end{table}

\begin{table}[bt]%The best place to locate the table environment is directly after its first reference in text
\caption{\label{4band}%
Parameters in the four-band effective model for monolayer FeTe with different lattice constants.}
%%%% new table
\begin{ruledtabular}
\begin{tabular}{cccccc}
%\textrm{Left\footnote{Note a.}}& \textrm{Centered\footnote{Note
%b.}}& \multicolumn{1}{c}{\textrm{Decimal}}&
%\textrm{Right}\\
%&\multicolumn{3}{c}{GaSe} & \multicolumn{3}{c}{GaS}\\
lattice(\AA) & $C$(eV)  & $M$(eV)  & $D$(eV\AA$^2$)  & $B$(eV\AA$^2$)  & $A$(eV\AA)  \\
 \colrule
3.805 & 0.082 & -0.221 & 3.381 & -4.122 & 2.318  \\
3.9 & 0.284 & -0.009 & -0.880 & -1.157 & 1.287 \\
3.925 & 0.336 & 0.045 & -0.586 & -0.100 & 1.426 \\
\end{tabular}
\end{ruledtabular}
\end{table}

\begin{figure}[t]
\centerline{\includegraphics[height=7 cm]{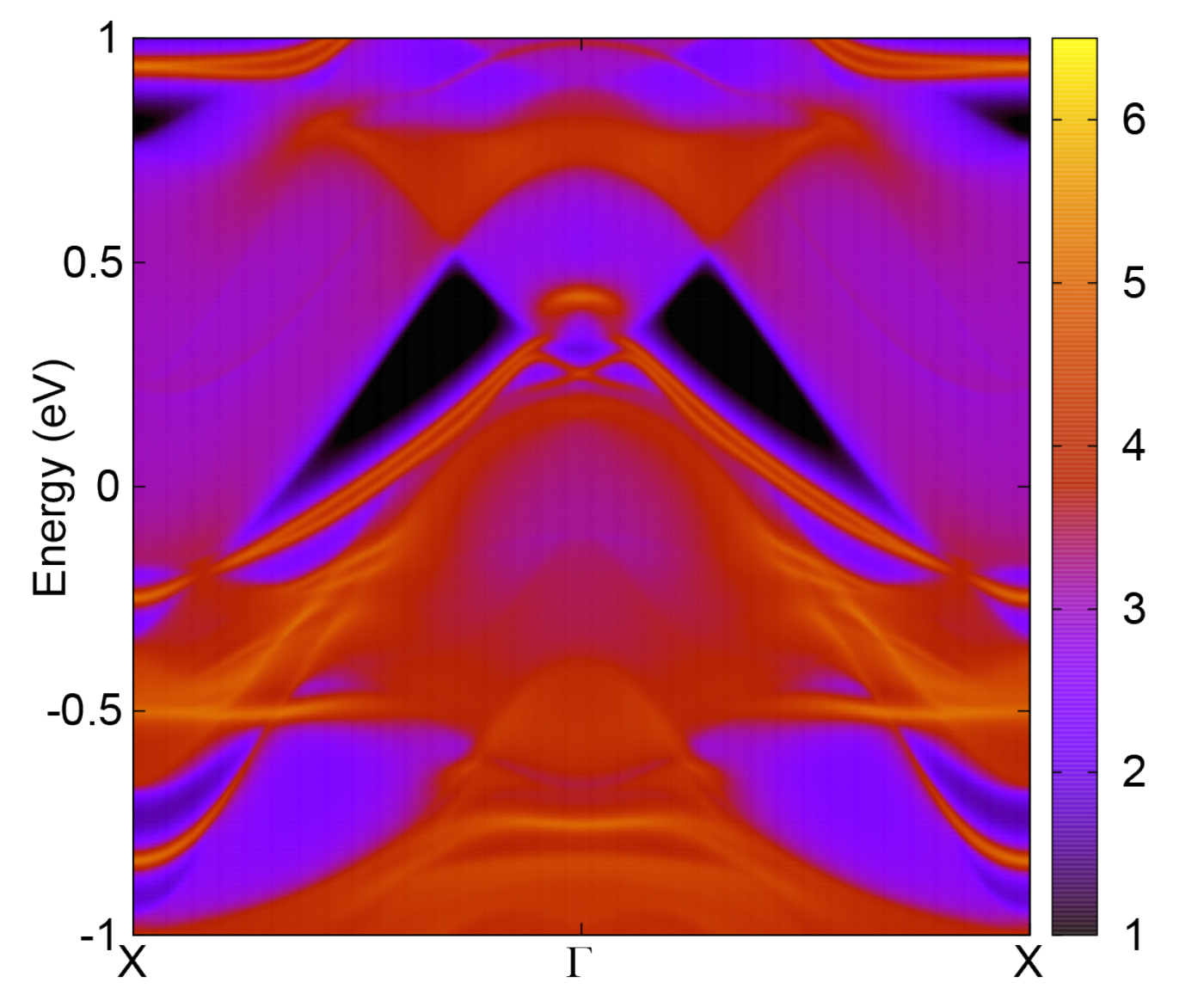}}
\caption{(color online). Energy and momentum dependence of the LDOS for the monolayer FeTe on the [100] edge. The higher LDOS is represented by brighter color.
 \label{edgestate} }
\end{figure}

As discussed above, the monolayer FeTe  is in the topologically  nontrivial region. To show the effect of nontrivial topology,  we calculate the surface Green's function of the semi-infinite system using an iterative method\cite{Sancho1984,Sancho1985}. The edge local density of states(LDOS) of [100] edge for FeTe monolayer
is shown in Fig.\ref{edgestate}. The nontrivial edge states confirm the conclusion that monolayer FeTe is topologically nontrivial by parity analysis.

{\it Multi-FeTe layers:}
The band structures of a bilayer and trilayer FeTe are given in supplementary material.  We find that while the bilayer system is topologically nontrivial, the trilayer system is topologically trivial.  A further calculation also shows that the quadruple-layer is also topologically nontrivial.  This oscillation behavior is similar to that of Bi$_2$Se$_3$ thin films\cite{Liu2010R} and can be argued as follows.  We can assume that there are only couplings in $A_{1u}$ states between two nearest-neighbour layers.   In a bilayer system,  a bonding and an anti-bonding bands are formed due to  the coupling between two layers. If the coupling is strong enough, the anti-bonding band gains energy and becomes unoccupied.  Therefore, the overall topological $Z_2$ character   is the same as in the single layer.  In a trilayer system, it is only one band among the  three reconstructed  bands due to the layer coupling  becomes unoccupied.  In this case, the overall $Z_2$ character changes and becomes trivial.  In a quadruple-layer systems, there are two bands among the four reconstructed bands  gain energy. However, one of them gains a much smaller energy. If this energy gain is not large enough to lift the band unoccupied,  the system has the non-trivial $Z_2$ topological invariance.  The argument can be quantified. Let $\Delta$ be the energy required to make the $A_{1u}$ band  unoccupied at $\Gamma$ point and $J$ is the layer coupling strength. To be consistent with our finding, we must have $0.61<\frac{\Delta}{J}<1$. According to our calculation, $\frac{\Delta}{J}$ is about 0.9, confirming the validity of our arguments.

% IFor a bilayer system, the inversion center is at in the middle of the two layers. Therefore, $\hat{I} c_{\alpha,u}(k)=c_{\alpha,b}(-k)$ and  $\hat{I} c_{\alpha,u}(k+Q)=-c_{\alpha,b}(-k+Q)$ for $\alpha=d$ orbitals and $\hat{I} c_{\alpha,u}(k)=-c_{\alpha,b}(-k)$ and  $\hat{I} c_{\alpha,u}(k+Q)=c_{\alpha,b}(-k+Q)$ for $\alpha=p$ orbitals, where $u$ and $b$ are the layer indices. Because of the interlayer coupling, the number of states near the Fermi level doubles compared with the case of monolayer FeTe. The antibonding states with odd parity have high energies. The $A_{1u}$ state contributed by $d_{xy}$ and $p_z$ have a strong interlayer coupling, leading to a large separation between $A^{-}_{1u}$ and $A^{+}_{1u}$. Similar to the monolayer case, the system is topological trivial if $A^{+}_{1u}$ and $A^{-}_{1u}$ states are above the $E^{-}_{g\pm}$ state. In bilayer FeTe with $a=3.809$\AA, $A^{+}_{1u}$ sinks below the $E^{-}_{g\pm}$ states. Now, the parity exchange has happened only once and the system is topological nontrivial. However, the parity exchange happens twice in trilayer FeTe, which drives the system into a topological trivial phase. It seems that parity exchange happens $N-1$ times in $N$ multi-layer FeTe ($N\geq2$). Therefore, the multi-layer FeTe is topologically nontrivial for even $N$ and topologically trivial for odd $N$. The $Z_2$ topological number is oscillating with the number of FeTe layers.

{\it Discussion:}
According to the aforementioned discussion, we know that the height of Te(Se) or lattice constants play an essential role in the topological phase transition.  Besides tuning the height  in FeTe$_{1-x}$Se$_x$  by changing $x$ to drive a topological phase transition,  it is known that  in the monolayer FeSe, the substrate can also affect the in-plane lattice constants and tune the Se height. For example, the measured lattice constant of FeSe on SrTiO$_3$ is about 3.82 \AA\cite{Zhang2014} while the lattice constant is 3.99 \AA ~ in FeSe/Nb:SrTiO$_3$/KTaO$_3$ heterostructures\cite{Peng2014}.  Therefore, the nontrivial topology discussed here may also exist in the monolayer FeSe on a substrate.
Moreover, the similar topological properties may exist  in iron pnictides.  Even if the strength of SOC in iron-pnictides is relatively weaker than those of iron-chalcogenides, the coupling between Fe and anions can still create  parity exchange. For bulk system, the parity exchange can happen at $Z$ point, which will drive the system into a strong topological phase.

The existence of nontrivial topology in FeTe$_{1-x}$Se$_x$  is supported  by the recent experimental observation\cite{Zhang2014_ding} in a bulk material.  Zhang {\it et~al.} observed an electron band above the Fermi level in Fe$Te_{0.55}$Se$_{0.45}$\cite{Zhang2014_ding}, which is rather similar to the bands of FeTe with $a=3.925$\AA.  It is also worth to mention that  a zero-bias peak (ZBP) was observed recently at an iron impurity site in the superconducting state of  FeTe$_{1-x}$Se$_x$\cite{Yin2014}. As the ZBP appears uniquely in this material,  it is  interesting to ask  whether it is related to the nontrivial topology.

In conclusion, we predict that the monolayer and the thin film    FeTe$_{1-x}$Se$_x$   is in a topological phase, which is induced by the parity exchange at $\Gamma$ point. The FeTe$_{1-x}$Se$_x$ can be  an ideal system for realizing topological superconductors and Majorana Fermions in a single phase. (We have learned that  an independent study on the bulk band structure  of FeTe$_{1-x}$Se$_x$   and its 3-dimensional nontrivial topology are being carried out simultaneously with this work.\cite{Wang2014})

We thank discussion with H. Ding. We thank X. Dai and Z. Fang for informing us their related independent work on bulk FeTe(Se).
The work is supported by the Ministry of Science and Technology of China 973
program(Grant No. 2012CV821400 and No. 2010CB922904), National Science Foundation of China (Grant No. NSFC-1190024, 11175248 and 11104339), and   the Strategic Priority Research Program of  CAS (Grant No. XDB07000000).

%\bibliographystyle{prsty}
%%\bibliographystyle{plain}
%%\bibliographystyle{unsrt}
%%\bibliographystyle{abbrv}
%%\bibliographystyle{alpha}
%\bibliography{cafeas2_ref}

\end{document}